\documentclass[12pt]{article}

\usepackage[left=1in,right=1in,top=1in,bottom=1in]{geometry}

\usepackage{setspace}
\usepackage{amsmath}
\usepackage{amsfonts}
\usepackage{comment}
\usepackage{float}
\usepackage{color}
\usepackage[font=small]{caption}
\usepackage{subcaption}
\usepackage{hyperref}
\hypersetup{colorlinks=true, citecolor=black, linkcolor=., urlcolor=cyan}
\usepackage[scr=boondox]{mathalfa}
\allowdisplaybreaks
\usepackage{natbib}
\usepackage{graphicx}
\graphicspath{{figures/}}
\usepackage{xcolor}

\newcommand{\rotX}{\tilde{\mathbf{X}}}
\newcommand{\roty}{\tilde{\mathbf{y}}}

\providecommand{\F}{\mathbf{F}}

\providecommand{\I}{\mathbf{I}}

\providecommand{\K}{\mathbf{K}}

\providecommand{\X}{\mathbf{X}}

\newcommand{\y}{\mathbf{y}}

\let\origc\c
\DeclareRobustCommand\c{\ifmmode\mathbf{c}\else\expandafter\origc\fi}
\let\origd\d
\DeclareRobustCommand\d{\ifmmode\mathbf{d}\else\expandafter\origd\fi}
\let\origu\u
\DeclareRobustCommand\u{\ifmmode\mathbf{u}\else\expandafter\origu\fi}
\let\origd\v
\DeclareRobustCommand\v{\ifmmode\mathbf{v}\else\expandafter\origv\fi}

\providecommand{\bb}{\boldsymbol{\beta}}

\providecommand{\bep}{\boldsymbol{\epsilon}}

\providecommand{\bS}{\boldsymbol{\Sigma}}

\providecommand{\zero}{\mathbf{0}}

\usepackage{booktabs}
\usepackage{multirow}
\usepackage{tabularx}
\newcolumntype{Y}{>{\centering\arraybackslash}X} 

\title{\vspace{-2.0em} plmmr: an R package to fit penalized linear mixed models for genome-wide association data with complex correlation structure}
\author{
  Tabitha K. Peter\\Dept. of Biostatistics\\University of Iowa
  \and
  Anna C. Reisetter\\Dept. of Biostatistics\\University of Iowa
  \and
  Yujing Lu\\Dept. of Biostatistics\\University of Iowa
  \and
  Oscar A. Rysavy\\
  Dept. of Biostatistics\\
  University of Iowa 
  \and 
  Patrick J. Breheny\\Dept. of Biostatistics\\University of Iowa
}

\begin{document}

\maketitle

\author{

}

\abstract{Correlation among the observations in high-dimensional regression modeling can be a major source of confounding. We present a new open-source package, \textbf{plmmr}, to implement \textbf{p}enalized \textbf{l}inear \textbf{m}ixed \textbf{m}odels in \textbf{R}. This R package estimates correlation among observations in high-dimensional data and uses those estimates to improve prediction with the best linear unbiased predictor. The package uses memory-mapping so that genome-scale data can be analyzed on ordinary machines even if the size of data exceeds RAM. We present here the methods, workflow, and file-backing approach upon which \textbf{plmmr} is built, and we demonstrate its computational capabilities with two examples from real GWAS data. 

\bigskip

}

\section{Background}\label{sec:intro}

Regression models for high-dimensional data have largely focused on independent observations, but correlation among samples can arise for many reasons, such as batch effects, geographic differences, ancestral groups, and/or family relationships. Such correlation can be a major source of confounding in data analysis. As a result, many approaches involve restricting the analysis to smaller groups of independent subjects. We present a new open-source package for high-dimensional regression capable of accounting for this correlation, thereby allowing the analysis to proceed incorporating data from all observations. Our package, \textbf{plmmr} (\url{https://github.com/pbreheny/plmmr}), implements \textbf{p}enalized \textbf{l}inear \textbf{m}ixed \textbf{m}odels in \textbf{R}. Of note, \textbf{plmmr} can handle latent/cryptic correlation structure (i.e., one does not need to know batch assignments or pedigree information), and scales up efficiently to handle genome-scale data such as genome-wide association studies (GWAS), even if the size of the data exceeds the memory of the machine.

Increasingly, batch effects have been recognized as having critical impacts on high-dimensional data \citep{Leek2010}. One approach to addressing this type of correlation is to derive additional covariates in the form of principal components (PCs) or surrogate variables (SVs) and include them in the analysis \citep{Price2006, Leek2007}, although there is an inherent challenge in determining how many PCs/SVs to include in the model. This type of correlation is increasingly common in the context of human genetics due to an emphasis on increasing the diversity in GWAS data by intentionally recruiting participants from other ancestry groups \citep{Mills2020}. Historically, most human genetics studies focused on homogeneous populations, with nearly 95\% of existing GWAS data representing people of European ancestry \citep{Mills2020}.

While batch-effects and population stratification result in large group structures, relational structures can also create small, highly-correlated groups. An important case of this is family-based studies in GWAS, which have been acknowledged as valuable for the field \citep{Benyamin2009}. At present, existing methodologies either assume that all family groups have the same known composition (e.g., all trios), or attempt to satisfy the assumption of independence by restricting the analysis to a set of unrelated individuals. However, identifying the largest subgroup of unrelated people in a given dataset is both an NP-hard problem and by definition results in excluding data from the analysis \citep{Galil1986, Toroslu2007, Abraham2014, Staples2013}.

Large-scale and small-scale relationships among observations are often present in the same data set, such as a GWAS containing family groups from different geographic regions. Such combinations of relationships result in complex correlation structure. Furthermore, it is typically unrealistic to assume full knowledge of this structure -- batch effects are usually not apparent, ancestry is complicated, and relationships may be cryptic. We describe in Section \ref{sec:preconditioning} the technique \textbf{plmmr} uses to accommodate complex correlation structures without requiring the relationships among observations to be known in advance.

An important distinction between \textbf{plmmr} and many other software packages that implement LMMs for high-dimensional data is that \textbf{plmmr} takes a joint modeling approach as opposed to a one-at-a-time (or `marginal') approach. A joint modeling approach is an additive model which considers the cumulative impact of all features in the data. A joint model identifies important features via sparsity-inducing penalties, such that the final model includes only the features that improve prediction of the outcome. As such, one advantage of the joint modeling approach over such a marginal approach is that in the former we directly construct a predictive model. This has implications for polygenic risk score calculation, as polygenic risk scores based on one-at-a-time testing require additional steps to combine multiple marginal models into a single prediction. Recognizing this advantage of joint modeling, several recent approaches (e.g., BOLT-LMM \citep{Loh2015}, SAIGE \citep{Zhou2018}, fastGWA \citep{fastGWA}, and REGENIE \citep{Mbatchou2021}) use a two-step approach in which a joint model is used as a first step. The joint modeling step is then followed by marginal testing designed to produce per-variant results. Our \textbf{plmmr} package offers something new as it implements a joint model in one single step -- results are provided from the regression model, instead of having a second step of marginal testing. 

Our presentation of the \textbf{plmmr} package is organized as follows: Section 2 summarizes the methodological approach for handling correlation, outlines the workflow of the \textbf{plmmr} pipeline, and describes the file-backing technique \textbf{plmmr} uses to scale up to large data. Section 3 presents computational time for \textbf{plmmr} analyses using real GWAS data. Finally, Section \ref{sec:discussion} situates \textbf{plmmr} in the current landscape of tools available for analyzing correlated GWAS data, outlining strengths, limitations, and future directions for our proposed approach.

\section{Implementation}

\subsection{Preconditioning a linear mixed model}\label{sec:preconditioning}
In order to incorporate complex correlation structure into the model for the data, \textbf{plmmr} uses a technique that projects the data onto a transformed scale. This technique has been called `preconditioning' in the literature -- for example, see \citet{Jia2015} or \citet{Wathen2015}. In brief, preconditioning requires a projection matrix (the `preconditioner') $\F$ and transforms the problem $\y = \X\bb$ into $\F\y = (\F\X)\bb$. 
In our model, we define $\X = n \times p$ as a standardized design matrix, and $\y = n \times 1$ as the outcome of interest. We then define $\K = \frac{1}{p}\X\X^\top$. Note that in the specific context of GWAS where $\X$ is a genotype matrix, $\K$ is known as the genomic relatedness matrix (GRM, also known as the ``kinship" matrix as defined by \citet{Thomas2005}). We then adopt the linear mixed model proposed by \citet{Rakitsch2013}:
\begin{equation}
\label{model}
   \y = \X\bb + \u + \bep
\end{equation}
where random effect $\u$ represents an \textbf{u}nobserved random effect with the distribution $\u \sim N(\zero, \sigma^2_s\K)$. Under the standard assumptions that $\bep \perp \u$ and $\bep \sim N(0, \sigma^2_\epsilon\I)$, the variance of $\y$ may be written $\bS = \sigma^2_s\K + \sigma^2_\epsilon\I$, with $\sigma^2_s$ representing the variance of $\y$ due to population \textbf{s}tructure and $\sigma^2_\epsilon$ represents the variation in $\y$ due to noise. Model \eqref{model} can therefore be equivalently written

\begin{equation}
    \label{plmmdist}
    \y \sim N(\X\bb, \sigma^2_s\K + \sigma^2_\epsilon\I) \equiv \y \sim N(\X\bb, \bS).
\end{equation}

We precondition equation \eqref{plmmdist}
using $\bS^{-1/2}$, to obtain
\begin{equation}
\label{precondmodel1}
\bS^{-1/2}\y \sim N((\bS^{-1/2}\X)\bb, \I),
\end{equation}
which we re-express as
\begin{equation}
\label{precondmodel2}
\roty \sim N(\rotX\bb, \I),
\end{equation}
where $\rotX$ and $\roty$ represent the design matrix and outcome vector on the rotated scale, respectively. As shown in Equation \ref{precondmodel2}, this preconditioning serves to `decorrelate' the variance structure so that observations on the $\roty$, $\rotX$ scale are independent. On this transformed scale, penalized regression approaches such as lasso \citep{Tibshirani1996}, SCAD \citep{Fan2001}, or MCP \citep{Zhang2010} may be applied \citep{Rakitsch2013, Jia2015, Cevid2020}.

\subsection{Workflow: from data files to model results}\label{sec:workflow}

With current available tools, carrying out the analysis described in \ref{sec:preconditioning} requires a variety of different software packages written in different languages. Users must link together these various tools, typically using command-line functions. Requiring each analyst to code their own pipeline is inefficient, error prone, and presents a barrier to reproducibility.

\begin{figure}[H]
    \centering
    \includegraphics[width=0.95\linewidth]{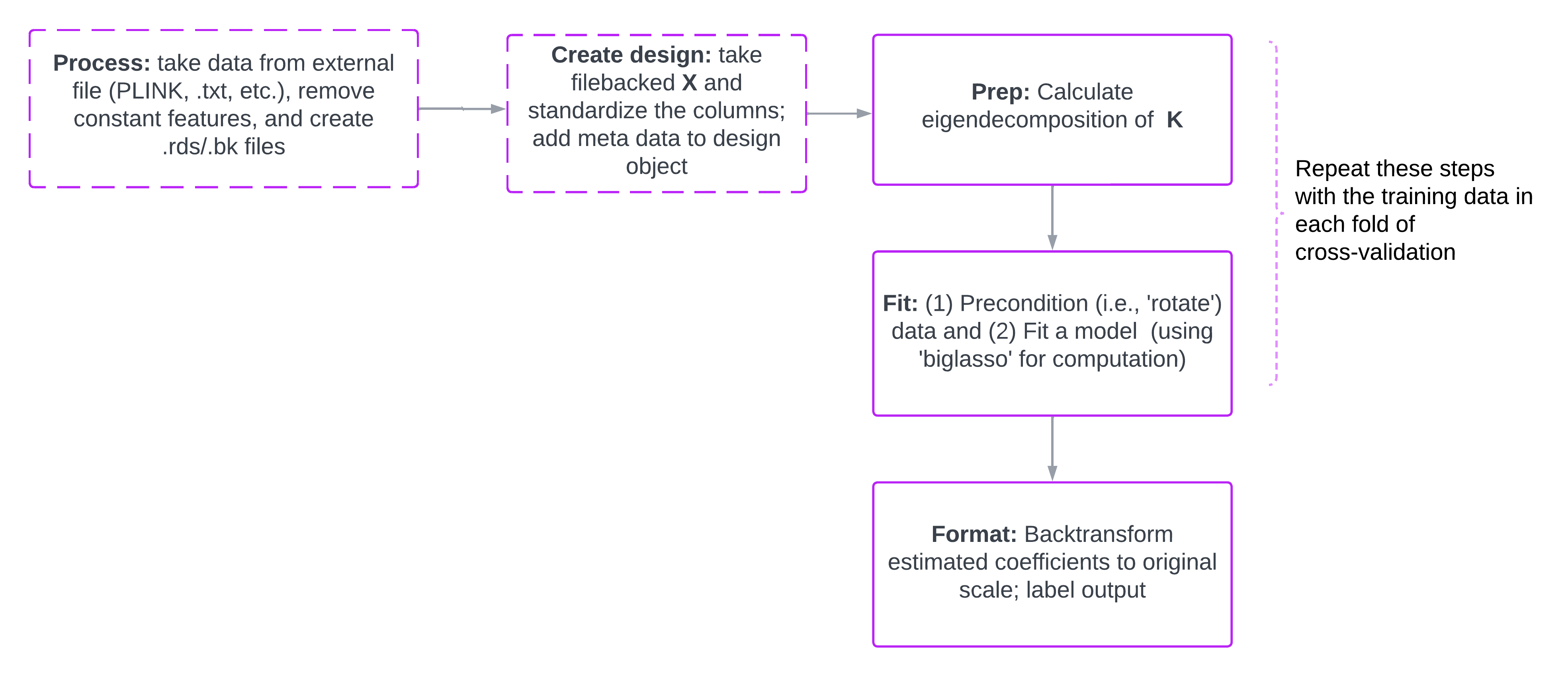}
    \caption{Workflow for plmmr. Steps shown with dotted lines are optional; steps shown with solid lines indicate essential components of the workflow.}
    \label{fig:workflow}
\end{figure}

This motivated us to create \textbf{plmmr}, which offers an integrated workflow as shown in Figure \ref{fig:workflow}. As an example of this workflow is shown in the R code below, which carries out a GWAS analysis in \textbf{plmmr} consisting of several steps: reading in PLINK files, estimating the relatedness matrix, preconditioning the data, fitting a model, and summarizing the results. Since all steps use the same R package, there is no need to convert between file types, data structures, programming languages, etc.

\begin{verbatim}
    # assuming that files plink.bed/plink.bim/plink.fam
    # are stored in directory "data_dir": 

    library(plmmr)
    # create filebacked object from PLINK data files 
    plink_data <- process_plink(data_dir = "data_dir", 
                                data_prefix = "plink",
                                rds_dir = "some_dir",
                                rds_prefix = "plink_data")

    # read in phenotype data 
    phen <- read.csv("clinical.csv") 

    # create a design
    design <- create_design(data_file = plink_data,
                            feature_id = "FID",
                            rds_dir = "some_dir",
                            new_file = "design",
                            add_outcome = phen,
                            outcome_id = "ID",
                            outcome_col = "outcome")
                            
    # fit a model 
    fit <- plmm(design) 

    # summarize coefficients at 50th lambda value
    summary(fit, idx = 50) 

    # plot of estimated coefficient paths
    plot(fit)  
\end{verbatim}

The first step in the workflow above involves creating an R object corresponding to the input data. \textbf{plmmr} is designed to accept multiple forms of data input, including a delimited file or a set of PLINK files. For data coming from external files too large to read into memory, the \textbf{plmmr} workflow includes a processing step that creates a pointer object to the external file(s) rather than reading them into R. This filebacking approach (described in greater detail in Section \ref{sec:filebacking}) allows \textbf{plmmr} to analyze GWAS-scale data even on machines where memory is limited.

Once there is an R object representing the data, \verb|create_design()| takes this object as input and implements the following measures to prepare data for model fitting: 

\begin{enumerate}
    \item Integrates outcome information 
    \item Option: designate additional, unpenalized features
    \item Standardizes design matrix $\X$ 
\end{enumerate}
Integrating the outcome information into the design is not necessarily trivial, as merging $\X$ and $\y$ requires proper alignment with respect to the order of the observations; \verb|create_design()| checks for this alignment, rather than assuming the user has already addressed this issue. The \verb|create_design()| function also has several options for designating unpenalized features; for GWAS data, features from another file such as age and sex may be merged in with the genotype data as unpenalized covariates in the model design. The final design matrix is column-standardized, and it is returned as part of the \verb|plmm_design| object returned by \verb|create_design()|. This object can be passed directly into the main model fitting function \verb|plmm()|. 

The internal work of the \verb|plmm()| function is made up of three steps, which we refer to as (1) the `prep' step, (2) the `fit' step, and (3) the `format' step. The `prep' step prepares the projection matrix to be used in analysis by taking an eigendecomposition of the matrix $\K$. The eigendecomposition of $\K$ is necessary for constructing the projection matrix $\bS^{-1/2}$. The `fit' step uses a coordinate descent algorithm to fit the model. \verb|plmm()| is designed to be flexible to the needs of the user, offering many optional arguments that allow the user to customize details such as the choice of $\lambda$ and the type of penalty (lasso/SCAD/MCP). The `format' step transforms the estimated coefficients back onto the scale of the original data -- this is done for clarity of interpretation. The results of \verb|plmm()| can be passed directly into \textbf{plmmr}'s \verb|plot()| and \verb|summary()| methods, so that there is seamless integration with simple syntax throughout the entire workflow. Figure \ref{fig:pofc_paths} is an example of the output from \verb|plot()|.

In addition to model fitting, \textbf{plmmr} also offers a cross-validation (CV) method, \verb|cv_plmm()|, that both fits a model and chooses its tuning parameter $\lambda$ with the syntax shown below: 
\begin{verbatim}
    cv_fit <- cv_plmm(design)

    # plot and summary methods: 
    summary(cv_fit)
    plot(cv_fit)
\end{verbatim}
Care must be taken when applying CV to the analytical approach of  \ref{sec:preconditioning}, as preconditioning has implications for exchangeability. Although standard penalized regression software can be used to fit a model on preconditioned data, the CV methods these other software supply will be incorrect if the preconditioning step is not included in each cross-validation fold. Correct implementation of CV requires that every part of the model fitting process be cross-validated \citep{Hastie2009}. A homebrewed pipeline is liable to get this part of the analysis wrong and lead to unintentional errors. We further developed these ideas in the methods work behind \textbf{plmmr}, so that the CV method in \textbf{plmmr} is integrated with the entire model-fitting process \citep{Rabinowicz2022a}. The \verb|cv_plmm()| return value may be passed directly to \verb|plot()| and \verb|summary()| methods as well; example output from \verb|plot()| is shown in Figure \ref{fig:pofc_cve}. 

\subsection{Prediction}\label{blup}

Depending on the scientific goal, prediction may be of equal or greater interest than identifying important features. Examples include predicting future clinical outcomes such as blood pressure and heart disease, making predictions in plant and animal breeding, developing polygenic risk scores, and inferring causal relationships using Mendelian randomization. Best linear unbiased prediction (BLUP) incorporates the correlation/relationship between outcomes in addition to the direct effects of individual features, and this approach increases accuracy in a wide variety of applications \citep{Robinson1991}. Since \textbf{plmmr} estimates the correlation among observations, it naturally lends itself to the use of BLUPs, which our package provides via the \verb|predict()| and \verb|cv_plmm()| functions. In other words, \textbf{plmmr} uses the estimated correlation structure not only to correct for potential confounding and reduce false positives, but also to improve prediction.

\subsection{Filebacking and Integration with C++}\label{sec:filebacking}

One major challenge in analyzing GWAS-scale data is the limitation of random-access memory (RAM), which motivated the design of \textbf{plmmr} as a package that uses filebacking. In cases where $\X$ is too large for one machine's RAM to accommodate, \textbf{plmmr} creates a file on disk, assigns a C++ pointer to this file, and allows that pointer to be accessible as an R object. The user then interacts with the pointer in the R session, so that the data are not read into memory. This technique of creating files on disk has been often employed to analyze large data \citep{Kane2013,Prive2018}. \textbf{plmmr} builds on the \textbf{bigmemory} package infrastructure for creating R objects that `point' to files on disk. The major model fitting steps use \textbf{bigalgebra} \citep{bigalgebra} and \textbf{biglasso} \citep{Zeng2021}, operating in C++ on the data stored in the binary file. This improves computational time and ensures that the design matrix, $\X$, is never read into memory. The output from a \verb|plmm()| model includes the estimated coefficients for each predictor at each value of the tuning parameter, saved in a sparse format as offered by the \textbf{Matrix} package \citep{Matrix}. In this way, the input to the model fitting function, the model fitting process itself, and the object returned are optimized to be memory-efficient and enable analyses to be run on a personal computer. 

\section{Results}

Computational time and contextualization with real data are paramount for ensuring that software is scaleable, accessible, and useful. In the following two examples, we use two real GWAS datasets to illustrate the performance of \textbf{plmmr}. Although the examples we present in Sections \ref{sec:penncath} and \ref{sec:pofc} focus on GWAS as the most computationally demanding type of analysis, we note that the \textbf{plmmr} package is also useful for other types of analysis beyond GWAS, such as gene expression analyses in the presence of possible batch effects. 

\subsection{Coronary artery disease GWAS}\label{sec:penncath}

\begin{figure}[H]
    \centering
    \includegraphics[width=0.75\linewidth]{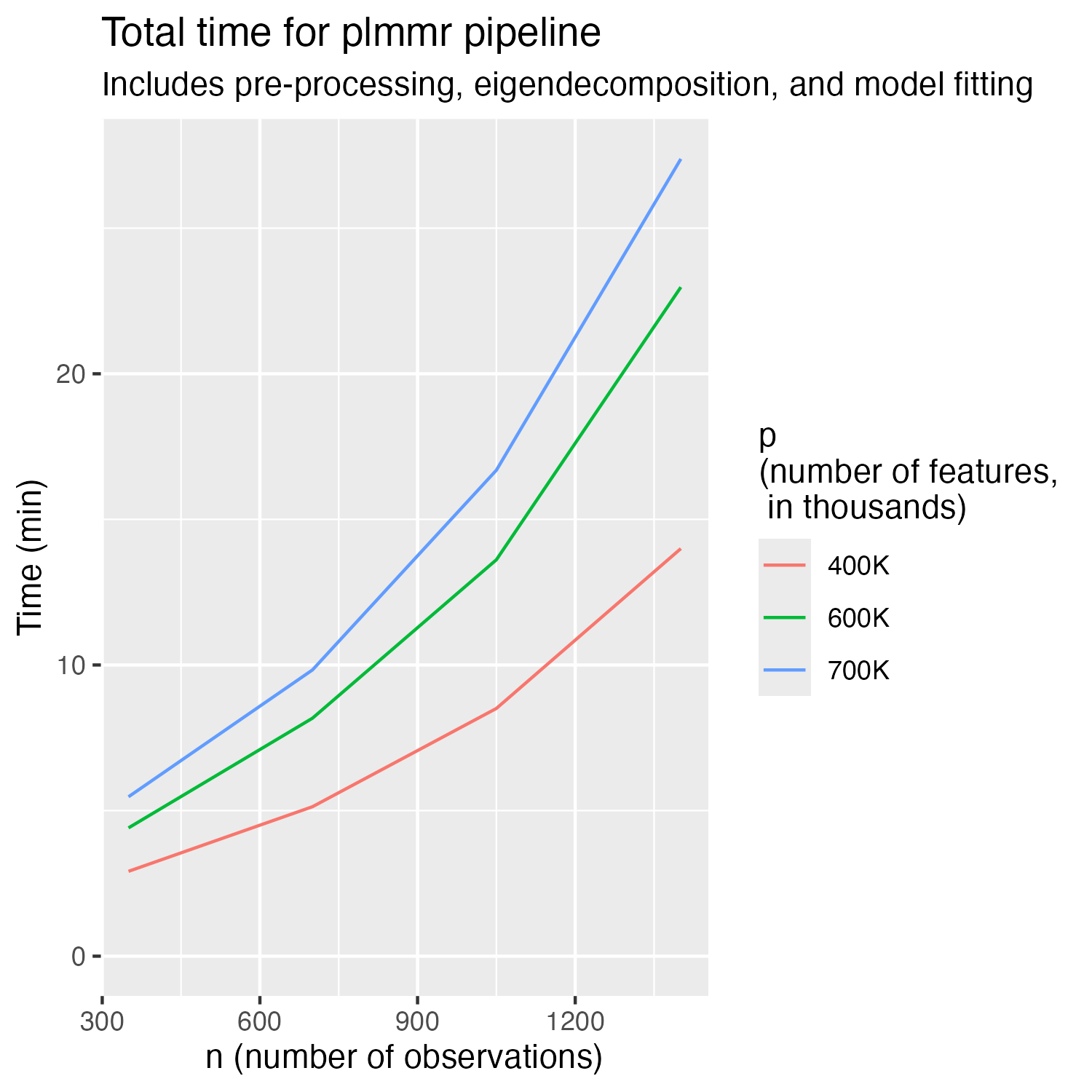}
    \caption{Total pipeline time}
    \label{fig:total-time}
\end{figure}

The PennCath study \citep{PennCath} was a population-based GWAS of 1,401 American participants of European ancestry in which the phenotype of interest was coronary artery disease. The genotype data for this study represent about 800,000 SNPs, and these data have been made publicly available. Starting with these data, we used genotype data from 696,644 autosomal SNPs that passed the quality control criteria (see Supplemental Material for quality control details) in order to illustrate the computational capabilities of \textbf{plmmr}. After quality control measures were taken, we created eleven subsets of genotype data using arbitrary filtering of samples and features. We varied the number of samples, $n$, so that $n \in \{350, 700, 1050, 1401\}$. We also varied the number of features, $p$, so that $p \in \{400\text{K}, 600\text{K}, 700\text{K}\}$ (where $\text{K} \equiv 1,000$). Each of our subsets of the PennCath data reflected one combination of these $n$ and $p$ values. For every data subset, we timed each step of the \textbf{plmmr} pipeline: reading in the PLINK files with \verb|process_plink()|, creating the design matrix with \verb|create_design()|, and fitting a penalized linear mixed model with \verb|plmm()|. Figure \ref{fig:total-time} illustrates the total time needed for the \textbf{plmmr} pipeline to fit a model on a laptop using a single core. 

\begin{figure}[H]
    \centering
    \includegraphics[width=0.75\linewidth]{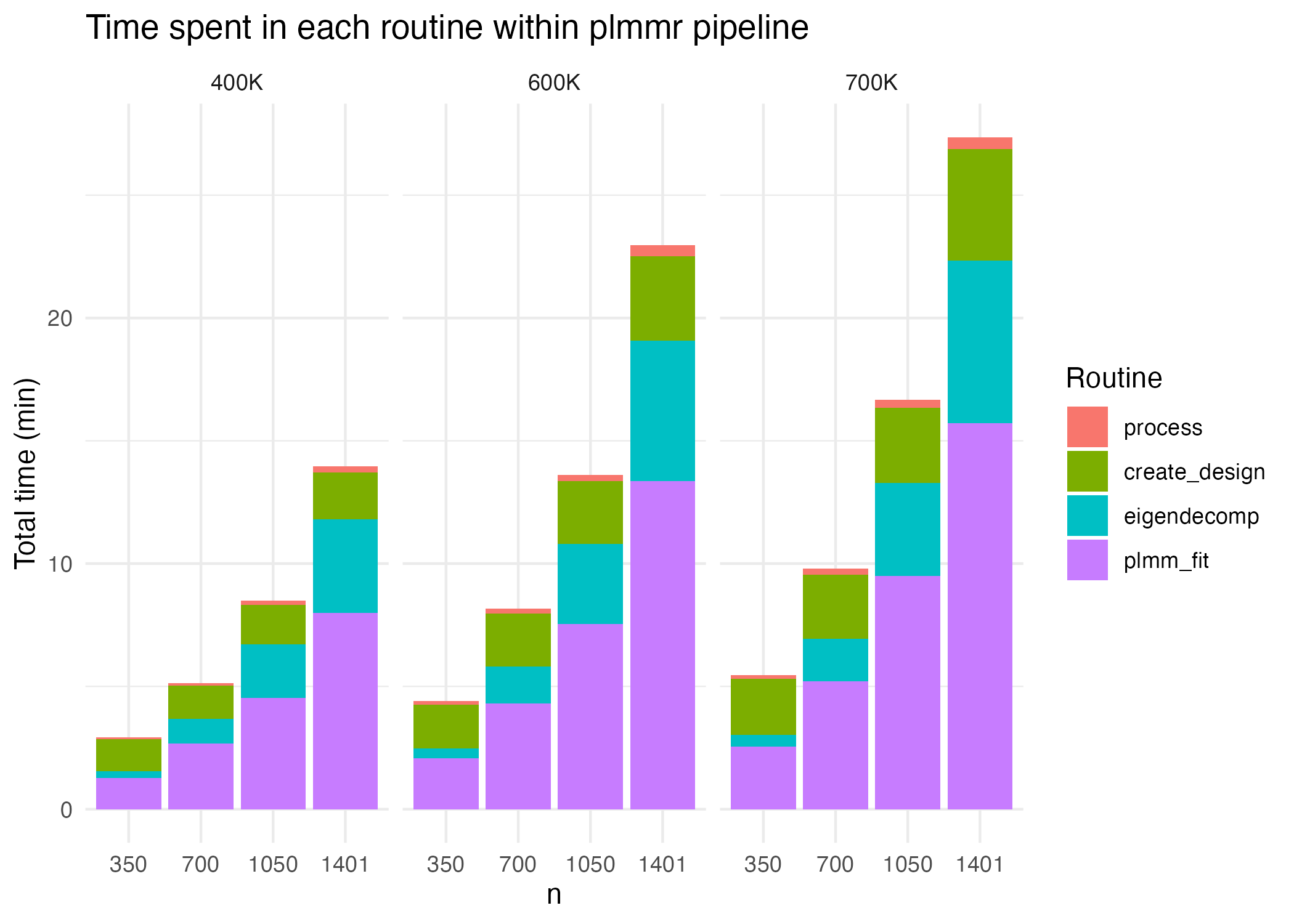}
    \caption{Time spent in each stage of pipeline}
    \label{fig:prop-breakdown}
\end{figure}

As summarized in Figure \ref{fig:prop-breakdown}, our results showed that model fitting time ranged from 1.5 minutes for the smallest subset ($n = 350, p = 400 \text{K}$) to 22 minutes for the full PennCath data ($n = 1,401, p = 700\text{K}$). We found that the pre-processing steps combined never took longer than about 5 minutes. We noticed that the increased computational time needed for larger values of $n$ was most attributable to the eigendecomposition of the realized relatedness matrix $\K$. Note that cross-validation would not necessarily require increased computational time, as CV can be parallelized. 

\subsection{Orofacial clefting GWAS}\label{sec:pofc}

To illustrate \textbf{plmmr} at work with more complex correlation structures, we used the \textbf{plmmr} pipleline to analyze data from the Pittsburgh Orofacial Cleft (POFC) study \citep{POFC2024} as our second example. The POFC study was a global, family-based GWAS in which the phenotype of focus was orofacial cleft (e.g., cleft palate). The GWAS data from the POFC study represents over 10,000 participants from over 2,500 families, and these families were recruited from fourteen global sites across five continents. The design matrix for this example included biological sex and country of recruitment site as unpenalized covariates, as these factors are known to be related to orofacial cleft formation \citep{Leslie2013}. While these genetic data were collected over ten years ago, \textbf{plmmr} has made it possible to include all of these participants (cleft patients, control patients, and all family members) in a single analysis for the first time.

\begin{figure}[H]
    \centering
    \includegraphics[width=0.75\linewidth]{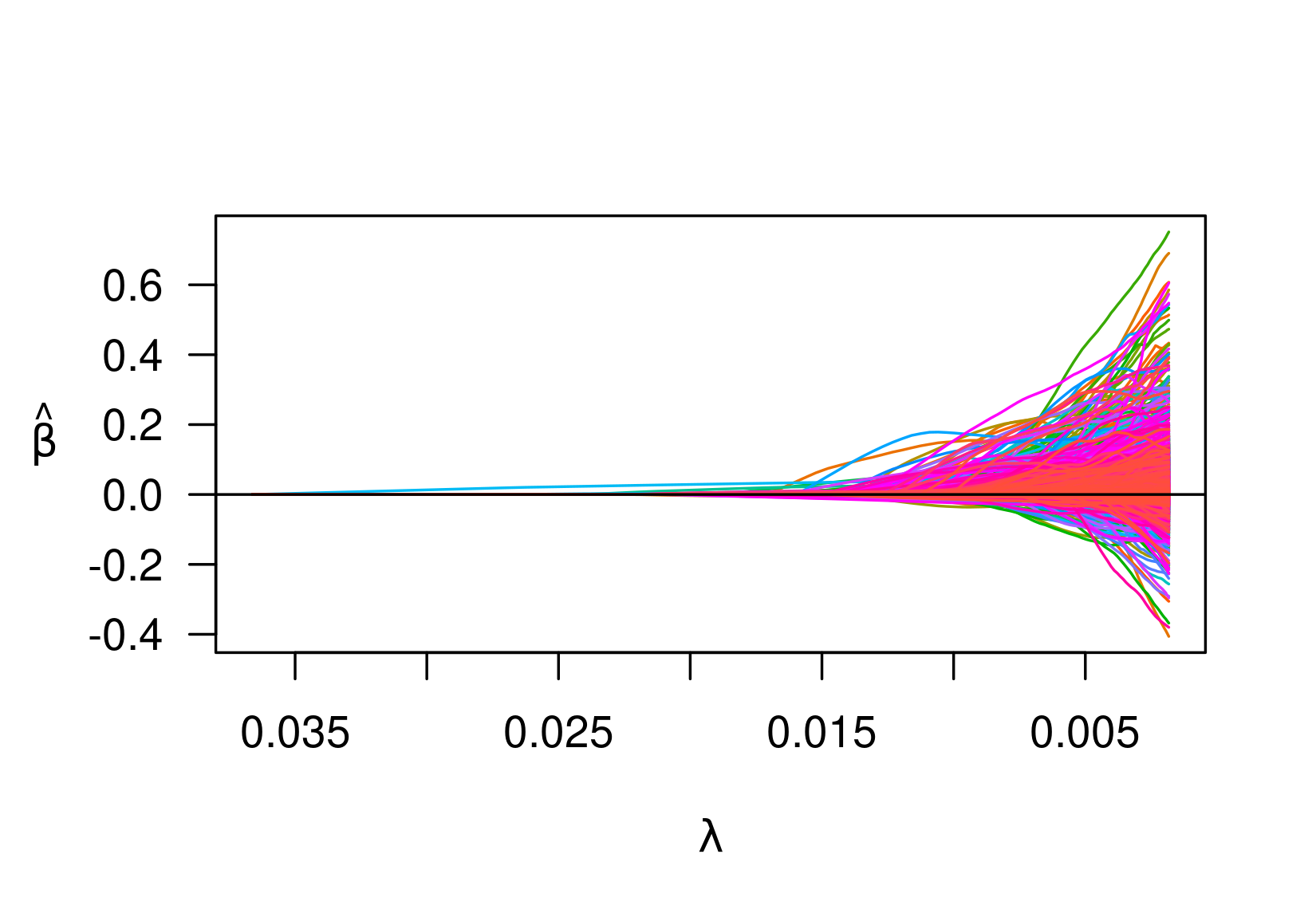}
    \caption{Plot of coefficient paths, POFC data}
    \label{fig:pofc_paths}
\end{figure}

\begin{figure}[H]
    \centering
    \includegraphics[width=0.75\linewidth]{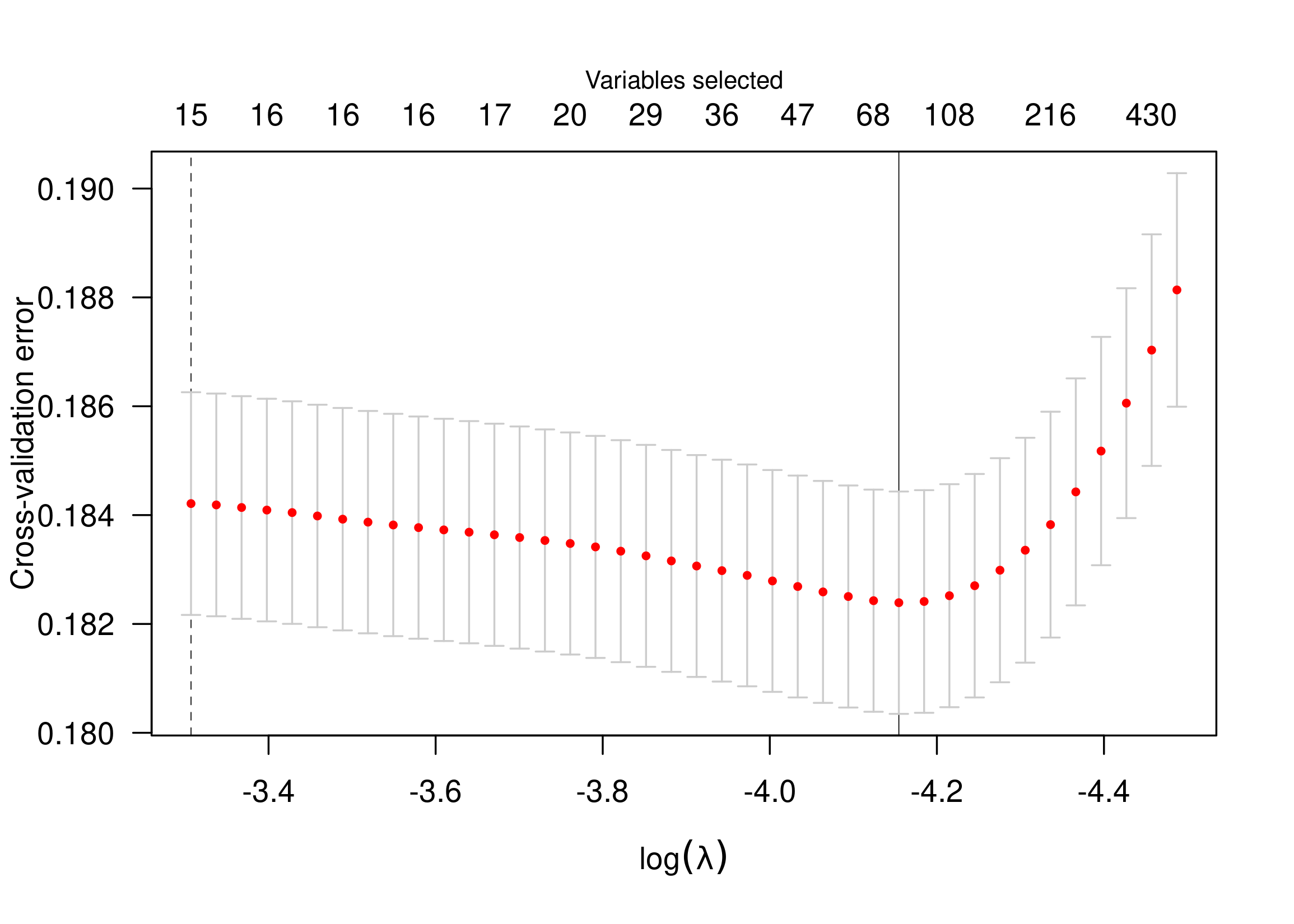}
    \caption{Plot of cross-validation error, POFC data \\(first 40 values of $\lambda$ shown)}
    \label{fig:pofc_cve}
\end{figure}

The POFC GWAS data represented 10,545 participants (samples), and 469,577 SNPs remained in the analytical data set after quality control measures were applied.
Due to the memory requirements of storing $\K$, an $n \times n$ matrix, with $n=10,545$, this analysis was run on a high-memory machine with an Intel Xeon CPU @ 2.40GHz processor. Creating the .rds and .bk files with \verb|plmmr::create_design()| took about nine minutes, and the eigendecomposition step took 15.4 hours. After the eigendecomposition step, the model fitting procedure required another 16.2 hours to complete. The selection of lasso tuning parameter $\lambda$ was done with 5-fold cross validation. Figure \ref{fig:pofc_cve} shows the cross-validation error (CVE) across the first 40 candidate values of $\lambda$, and Figure \ref{fig:pofc_paths} illustrates the estimated coefficient paths. At the $\lambda$ value which minimized cross-validation error, the lasso model selected 53 SNPs as having non-zero coefficients. The genes represented by these selected SNPs included several genes that have been identified as associated with orofacial clefts in previous literature: NTN1, PAX7, IRF6, and FOXE1 \citep{Leslie2015a, Leslie2015, Beaty2016}. 

\section{Discussion}\label{sec:discussion}

The \textbf{plmmr} package implements a joint mixed modeling approach for selecting features of interest while accounting for correlation. Several related tools have also been developed. Both the \textbf{glmmPen} \citep{glmmPen} and \textbf{HighDimMixedModels} \citep{Gorstein2024} packages implement penalized mixed models, but these packages assume that the factors which govern the correlation between subjects are known, and cannot be directly applied to the setting in which these relationships must be inferred or estimated, as in population genetics.
Other packages in the literature of joint mixed models for correlated, high-dimensional data include the R package \textbf{ggmix} \citep{ggmix}. While \textbf{ggmix} uses a similar transformation technique as \textbf{plmmr}, \textbf{ggmix} does not scale up to large, genome-scale data as well as \textbf{plmmr}, both in terms of speed and in terms of the capability to fit data larger than memory. In addition, the package does not currently offer a cross-validation method. \textbf{plmmr} is the only R package we know of that offers a file-backed, fully-integrated workflow that includes BLUP-based CV. Alongside this integrated workflow, \textbf{plmmr} offers thorough documentation including vignettes that users can work through using the datasets that ship with the package. This documentation gives \textbf{plmmr} an accessibility that is not common among bioinformatics softwares.  

One limitation of \textbf{plmmr} is that the required eigendecomposition of genomic relatedness matrix $\K$ is computationally expensive when $n$ is large. Figure~\ref{fig:total-time} illustrates that computational time does not scale linearly with $n$; indeed, the eigendecomposition of $\K$ scales with $n^2$. This is also reflected in the computational time needed for the POFC data example in Section~\ref{sec:pofc}. One potential approach to improving scalability is to adopt a hybrid perspective, combining a sparse $\K$ with principal components as unpenalized covariates as proposed in \citet{STAAR}.

Another limitation of \textbf{plmmr} is that it currently does not offer logistic regression. Binary outcomes can be analyzed, but they must be treated as numeric and analyzed with linear models. We are actively working to extend the penalized linear mixed modeling framework presented here to include logistic regression. A Julia package, \textbf{PenalizedGLMM} \citep{PenalizedGLMM}, was recently developed and fills an important gap in this area with its support for logistic regression. Our early experience indicates that \textbf{plmmr} is more efficient for fitting linear regression models, although \textbf{PenalizedGLMM}'s logistic regression functionality make the two packages useful, complementary tools for different modeling needs.

\section{Conclusions}

We have presented here a new R package, \textbf{plmmr}, which offers the capacity to fit penalized linear mixed models to GWAS-scale data with complex correlation structure. The software provides an end-to-end workflow that takes the user through all steps of the analysis in a single integrated pipeline, from processing raw data (e.g., PLINK files) to model summaries. 

\bibliographystyle{plainnat}

\begin{thebibliography}{42}
\providecommand{\natexlab}[1]{#1}
\providecommand{\url}[1]{\texttt{#1}}
\expandafter\ifx\csname urlstyle\endcsname\relax
  \providecommand{\doi}[1]{doi: #1}\else
  \providecommand{\doi}{doi: \begingroup \urlstyle{rm}\Url}\fi

\bibitem[Abraham and Diaz(2014)]{Abraham2014}
Kuruvilla~Joseph Abraham and Clara Diaz.
\newblock Identifying large sets of unrelated individuals and unrelated
  markers.
\newblock \emph{Source code for biology and medicine}, 9:\penalty0 1--8, 2014.

\bibitem[Bates et~al.(2024)Bates, Maechler, and Jagan]{Matrix}
Douglas Bates, Martin Maechler, and Mikael Jagan.
\newblock \emph{Matrix: Sparse and Dense Matrix Classes and Methods}, 2024.
\newblock URL \url{https://CRAN.R-project.org/package=Matrix}.
\newblock R package version 1.7-0.

\bibitem[Beaty et~al.(2016)Beaty, Marazita, and Leslie]{Beaty2016}
Terri~H Beaty, Mary~L Marazita, and Elizabeth~J Leslie.
\newblock Genetic factors influencing risk to orofacial clefts: today’s
  challenges and tomorrow’s opportunities.
\newblock \emph{F1000Research}, 5, 2016.

\bibitem[Benyamin et~al.(2009)Benyamin, Visscher, and McRae]{Benyamin2009}
Beben Benyamin, Peter~M Visscher, and Allan~F McRae.
\newblock Family-based genome-wide association studies.
\newblock \emph{Pharmacogenomics}, 10\penalty0 (2):\penalty0 181--190, 2009.

\bibitem[Bertrand et~al.(2024)Bertrand, Kane, Emerson, and Weston]{bigalgebra}
Frederic Bertrand, Michael~J. Kane, John Emerson, and Stephen Weston.
\newblock \emph{'BLAS' and 'LAPACK' Routines for Native R Matrices and
  'big.matrix' Objects}, 2024.
\newblock URL \url{https://fbertran.github.io/bigalgebra/}.
\newblock R package version 1.1.2.

\bibitem[Bhatnagar et~al.(2020)Bhatnagar, Yang, Lu, Schurr, Loredo-Osti,
  Forest, Oualkacha, and Greenwood]{ggmix}
Sahir~R Bhatnagar, Yi~Yang, Tianyuan Lu, Erwin Schurr, JC~Loredo-Osti, Marie
  Forest, Karim Oualkacha, and Celia~MT Greenwood.
\newblock Simultaneous snp selection and adjustment for population structure in
  high dimensional prediction models.
\newblock \emph{PLoS genetics}, 16\penalty0 (5):\penalty0 e1008766, 2020.

\bibitem[{\'C}evid et~al.(2020){\'C}evid, B{\"u}hlmann, and
  Meinshausen]{Cevid2020}
Domagoj {\'C}evid, Peter B{\"u}hlmann, and Nicolai Meinshausen.
\newblock Spectral deconfounding via perturbed sparse linear models.
\newblock \emph{Journal of Machine Learning Research}, 21\penalty0
  (232):\penalty0 1--41, 2020.

\bibitem[Fan and Li(2001)]{Fan2001}
Jianqing Fan and Runze Li.
\newblock Variable selection via nonconcave penalized likelihood and its oracle
  properties.
\newblock \emph{Journal of the American statistical Association}, 96\penalty0
  (456):\penalty0 1348--1360, 2001.

\bibitem[Galil(1986)]{Galil1986}
Zvi Galil.
\newblock Efficient algorithms for finding maximum matching in graphs.
\newblock \emph{ACM Computing Surveys (CSUR)}, 18\penalty0 (1):\penalty0
  23--38, 1986.

\bibitem[Gorstein et~al.(2024)Gorstein, Aghdam, and
  Sol'{i}s-Lemus]{Gorstein2024}
E.~Gorstein, R.~Aghdam, and C.~Sol'{i}s-Lemus.
\newblock {HighDimMixedModels.jl: Robust High Dimensional Mixed Models across
  Omics Data}.
\newblock \emph{In preparation}, 2024.

\bibitem[Hastie et~al.(2009)Hastie, Tibshirani, Friedman, and
  Friedman]{Hastie2009}
Trevor Hastie, Robert Tibshirani, Jerome~H Friedman, and Jerome~H Friedman.
\newblock \emph{The elements of statistical learning: data mining, inference,
  and prediction}, volume~2.
\newblock Springer, 2009.

\bibitem[Heiling et~al.(2024)Heiling, Rashid, Li, and Ibrahim]{glmmPen}
Hillary~M. Heiling, Naim~U. Rashid, Quefeng Li, and Joseph~G. Ibrahim.
\newblock glmmpen: High dimensional penalized generalized linear mixed models.
\newblock \emph{The R Journal}, 15:\penalty0 106--128, 2024.
\newblock ISSN 2073-4859.
\newblock \doi{10.32614/RJ-2023-086}.
\newblock https://doi.org/10.32614/RJ-2023-086.

\bibitem[Jia and Rohe(2015)]{Jia2015}
Jinzhu Jia and Karl Rohe.
\newblock Preconditioning the lasso for sign consistency.
\newblock \emph{Electronic Journal of Statistics}, 9\penalty0 (1):\penalty0
  1150--1172, 2015.
\newblock \doi{10.1214/15-EJS1029}.

\bibitem[Jiang et~al.(2019)Jiang, Zheng, Qi, Kemper, Wray, Visscher, and
  Yang]{fastGWA}
Longda Jiang, Zhili Zheng, Ting Qi, Kathryn~E Kemper, Naomi~R Wray, Peter~M
  Visscher, and Jian Yang.
\newblock A resource-efficient tool for mixed model association analysis of
  large-scale data.
\newblock \emph{Nature genetics}, 51\penalty0 (12):\penalty0 1749--1755, 2019.

\bibitem[Kane et~al.(2013)Kane, Emerson, and Weston]{Kane2013}
Michael~J. Kane, John~W. Emerson, and Stephen Weston.
\newblock Scalable strategies for computing with massive data.
\newblock \emph{Journal of Statistical Software}, 55\penalty0 (14):\penalty0
  1--19, 2013.
\newblock URL \url{https://www.jstatsoft.org/article/view/v055i14}.

\bibitem[Leek and Storey(2007)]{Leek2007}
J.~T. Leek and J.~D. Storey.
\newblock Capturing heterogeneity in gene expression studies by surrogate
  variable analysis.
\newblock \emph{PLoS Genetics}, 3\penalty0 (9):\penalty0 e161, 2007.
\newblock \doi{10.1371/journal.pgen.0030161}.

\bibitem[Leek et~al.(2010)Leek, Scharpf, Bravo, Simcha, Langmead, Johnson,
  Geman, Baggerly, and Irizarry]{Leek2010}
Jeffrey~T. Leek, Robert~B. Scharpf, H{\'e}ctor~Corrada Bravo, David Simcha,
  Benjamin Langmead, W.~Evan Johnson, Donald Geman, Keith Baggerly, and
  Rafael~A. Irizarry.
\newblock Tackling the widespread and critical impact of batch effects in
  high-throughput data.
\newblock \emph{Nature Reviews Genetics}, 11\penalty0 (10):\penalty0 733--739,
  October 2010.
\newblock ISSN 1471-0056.
\newblock \doi{10.1038/nrg2825}.

\bibitem[Leslie et~al.(2015{\natexlab{a}})Leslie, Koboldt, Kang, Ma, Hecht,
  Wehby, Christensen, Czeizel, Deleyiannis, Fulton, Wilson, Beaty, Schutte,
  Murray, and Marazita]{Leslie2015a}
E.J. Leslie, D.C. Koboldt, C.J. Kang, L.~Ma, J.T. Hecht, G.L. Wehby,
  K.~Christensen, A.E. Czeizel, F.W.-B. Deleyiannis, R.S. Fulton, R.K. Wilson,
  T.H. Beaty, B.C. Schutte, J.C. Murray, and M.L. Marazita.
\newblock {IRF}6mutation screening in non-syndromic orofacial clefting:
  analysis of 1521 families.
\newblock \emph{Clinical Genetics}, 90\penalty0 (1):\penalty0 28--34, oct
  2015{\natexlab{a}}.
\newblock \doi{10.1111/cge.12675}.

\bibitem[Leslie and Marazita(2013)]{Leslie2013}
Elizabeth~J Leslie and Mary~L Marazita.
\newblock Genetics of cleft lip and cleft palate.
\newblock \emph{American Journal of Medical Genetics Part C: Seminars in
  Medical Genetics}, 163\penalty0 (4):\penalty0 246--258, 2013.
\newblock \doi{https://doi.org/10.1002/ajmg.c.31381}.

\bibitem[Leslie et~al.(2015{\natexlab{b}})Leslie, Taub, Liu, Steinberg,
  Koboldt, Zhang, Carlson, Hetmanski, Wang, Larson, Fulton, Kousa, Fakhouri,
  Naji, Ruczinski, Begum, Parker, Busch, Standley, Rigdon, Hecht, Scott, Wehby,
  Christensen, Czeizel, Deleyiannis, Schutte, Wilson, Cornell, Lidral,
  Weinstock, Beaty, Marazita, and Murray]{Leslie2015}
Elizabeth~J. Leslie, Margaret~A. Taub, Huan Liu, Karyn~Meltz Steinberg,
  Daniel~C. Koboldt, Qunyuan Zhang, Jenna~C. Carlson, Jacqueline~B. Hetmanski,
  Hang Wang, David~E. Larson, Robert~S. Fulton, Youssef~A. Kousa, Walid~D.
  Fakhouri, Ali Naji, Ingo Ruczinski, Ferdouse Begum, Margaret~M. Parker,
  Tamara Busch, Jennifer Standley, Jennifer Rigdon, Jacqueline~T. Hecht,
  Alan~F. Scott, George~L. Wehby, Kaare Christensen, Andrew~E. Czeizel,
  Frederic~W.-B. Deleyiannis, Brian~C. Schutte, Richard~K. Wilson, Robert~A.
  Cornell, Andrew~C. Lidral, George~M. Weinstock, Terri~H. Beaty, Mary~L.
  Marazita, and Jeffrey~C. Murray.
\newblock Identification of functional variants for cleft lip with or without
  cleft palate in or near {PAX}7, {FGFR}2, and {NOG} by targeted sequencing of
  {GWAS} loci.
\newblock \emph{The American Journal of Human Genetics}, 96\penalty0
  (3):\penalty0 397--411, mar 2015{\natexlab{b}}.
\newblock \doi{10.1016/j.ajhg.2015.01.004}.

\bibitem[Li et~al.(2020)Li, Li, Zhou, Gaynor, Liu, Chen, Sun, Dey, Arnett,
  Aslibekyan, et~al.]{STAAR}
Xihao Li, Zilin Li, Hufeng Zhou, Sheila~M Gaynor, Yaowu Liu, Han Chen, Ryan
  Sun, Rounak Dey, Donna~K Arnett, Stella Aslibekyan, et~al.
\newblock Dynamic incorporation of multiple in silico functional annotations
  empowers rare variant association analysis of large whole-genome sequencing
  studies at scale.
\newblock \emph{Nature genetics}, 52\penalty0 (9):\penalty0 969--983, 2020.

\bibitem[Loh et~al.(2015)Loh, Tucker, Bulik-Sullivan, Vilhjalmsson, Finucane,
  Salem, Chasman, Ridker, Neale, Berger, et~al.]{Loh2015}
Po-Ru Loh, George Tucker, Brendan~K Bulik-Sullivan, Bjarni~J Vilhjalmsson,
  Hilary~K Finucane, Rany~M Salem, Daniel~I Chasman, Paul~M Ridker, Benjamin~M
  Neale, Bonnie Berger, et~al.
\newblock Efficient bayesian mixed-model analysis increases association power
  in large cohorts.
\newblock \emph{Nature genetics}, 47\penalty0 (3):\penalty0 284--290, 2015.

\bibitem[Marazita and Weinberg(2024)]{POFC2024}
Mary Marazita and Seth Weinberg.
\newblock Pittsburgh orofacial cleft studies, September 2024.
\newblock URL
  \url{https://www.dental.pitt.edu/research/ccdg/participate-research/pittsburgh-orofacial-cleft-studies}.
\newblock Center for Craniofacial and Dental Genetics, University of
  Pittsburgh. Website.

\bibitem[Mbatchou et~al.(2021)Mbatchou, Barnard, Backman, Marcketta, Kosmicki,
  Ziyatdinov, Benner, O’Dushlaine, Barber, Boutkov, et~al.]{Mbatchou2021}
Joelle Mbatchou, Leland Barnard, Joshua Backman, Anthony Marcketta, Jack~A
  Kosmicki, Andrey Ziyatdinov, Christian Benner, Colm O’Dushlaine, Mathew
  Barber, Boris Boutkov, et~al.
\newblock Computationally efficient whole-genome regression for quantitative
  and binary traits.
\newblock \emph{Nature genetics}, 53\penalty0 (7):\penalty0 1097--1103, 2021.

\bibitem[Mills and Rahal(2020)]{Mills2020}
Melinda~C Mills and Charles Rahal.
\newblock The gwas diversity monitor tracks diversity by disease in real time.
\newblock \emph{Nature genetics}, 52\penalty0 (3):\penalty0 242--243, 2020.

\bibitem[Price et~al.(2006)Price, Patterson, Plenge, Weinblatt, Shadick, and
  Reich]{Price2006}
Alkes~L Price, Nick~J Patterson, Robert~M Plenge, Michael~E Weinblatt, Nancy~A
  Shadick, and David Reich.
\newblock Principal components analysis corrects for stratification in
  genome-wide association studies.
\newblock \emph{Nature genetics}, 38\penalty0 (8):\penalty0 904--909, 2006.

\bibitem[Priv{\'e} et~al.(2018)Priv{\'e}, Aschard, Ziyatdinov, and
  Blum]{Prive2018}
Florian Priv{\'e}, Hugues Aschard, Andrey Ziyatdinov, and Michael~GB Blum.
\newblock Efficient analysis of large-scale genome-wide data with two r
  packages: bigstatsr and bigsnpr.
\newblock \emph{Bioinformatics}, 34\penalty0 (16):\penalty0 2781--2787, 2018.

\bibitem[Purcell et~al.(2007)Purcell, Neale, Todd-Brown, Thomas, Ferreira,
  Bender, Maller, Sklar, De~Bakker, Daly, et~al.]{plink}
Shaun Purcell, Benjamin Neale, Kathe Todd-Brown, Lori Thomas, Manuel~AR
  Ferreira, David Bender, Julian Maller, Pamela Sklar, Paul~IW De~Bakker,
  Mark~J Daly, et~al.
\newblock Plink: a tool set for whole-genome association and population-based
  linkage analyses.
\newblock \emph{The American journal of human genetics}, 81\penalty0
  (3):\penalty0 559--575, 2007.

\bibitem[{R Core Team}(2024)]{R}
{R Core Team}.
\newblock \emph{R: A Language and Environment for Statistical Computing}.
\newblock R Foundation for Statistical Computing, Vienna, Austria, 2024.
\newblock URL \url{https://www.R-project.org/}.

\bibitem[Rabinowicz and Rosset(2022)]{Rabinowicz2022a}
Assaf Rabinowicz and Saharon Rosset.
\newblock Cross-validation for correlated data.
\newblock \emph{Journal of the American Statistical Association}, 117\penalty0
  (538):\penalty0 718--731, 2022.

\bibitem[Rakitsch et~al.(2013)Rakitsch, Lippert, Stegle, and
  Borgwardt]{Rakitsch2013}
Barbara Rakitsch, Christoph Lippert, Oliver Stegle, and Karsten Borgwardt.
\newblock A lasso multi-marker mixed model for association mapping with
  population structure correction.
\newblock \emph{Bioinformatics}, 29\penalty0 (2):\penalty0 206--214, 2013.

\bibitem[Reilly et~al.(2011)Reilly, Li, He, Ferguson, Stylianou, Mehta,
  Burnett, Devaney, Knouff, Thompson, et~al.]{PennCath}
Muredach~P Reilly, Mingyao Li, Jing He, Jane~F Ferguson, Ioannis~M Stylianou,
  Nehal~N Mehta, Mary~Susan Burnett, Joseph~M Devaney, Christopher~W Knouff,
  John~R Thompson, et~al.
\newblock Identification of adamts7 as a novel locus for coronary
  atherosclerosis and association of abo with myocardial infarction in the
  presence of coronary atherosclerosis: two genome-wide association studies.
\newblock \emph{The Lancet}, 377\penalty0 (9763):\penalty0 383--392, 2011.

\bibitem[Robinson(1991)]{Robinson1991}
G.~K. Robinson.
\newblock That {BLUP} is a good thing: The estimation of random effects.
\newblock \emph{Statistical Science}, 6\penalty0 (1):\penalty0 15--32, 1991.

\bibitem[St-Pierre et~al.(2023)St-Pierre, Oualkacha, and
  Bhatnagar]{PenalizedGLMM}
Julien St-Pierre, Karim Oualkacha, and Sahir~Rai Bhatnagar.
\newblock Efficient penalized generalized linear mixed models for variable
  selection and genetic risk prediction in high-dimensional data.
\newblock \emph{Bioinformatics}, 39\penalty0 (2):\penalty0 btad063, 2023.

\bibitem[Staples et~al.(2013)Staples, Nickerson, and Below]{Staples2013}
Jeffrey Staples, Deborah~A Nickerson, and Jennifer~E Below.
\newblock Utilizing graph theory to select the largest set of unrelated
  individuals for genetic analysis.
\newblock \emph{Genetic epidemiology}, 37\penalty0 (2):\penalty0 136--141,
  2013.

\bibitem[Thomas(2005)]{Thomas2005}
Stuart~C Thomas.
\newblock The estimation of genetic relationships using molecular markers and
  their efficiency in estimating heritability in natural populations.
\newblock \emph{Philosophical Transactions of the Royal Society B: Biological
  Sciences}, 360\penalty0 (1459):\penalty0 1457--1467, 2005.

\bibitem[Tibshirani(1996)]{Tibshirani1996}
Robert Tibshirani.
\newblock Regression shrinkage and selection via the lasso.
\newblock \emph{Journal of the Royal Statistical Society Series B: Statistical
  Methodology}, 58\penalty0 (1):\penalty0 267--288, 1996.

\bibitem[Toroslu and Arslanoglu(2007)]{Toroslu2007}
Ismail~H Toroslu and Yilmaz Arslanoglu.
\newblock Genetic algorithm for the personnel assignment problem with multiple
  objectives.
\newblock \emph{Information Sciences}, 177\penalty0 (3):\penalty0 787--803,
  2007.

\bibitem[Wathen(2015)]{Wathen2015}
Andrew~J Wathen.
\newblock Preconditioning.
\newblock \emph{Acta Numerica}, 24:\penalty0 329--376, 2015.

\bibitem[Zeng and Breheny(2021)]{Zeng2021}
Yaohui Zeng and Patrick Breheny.
\newblock The biglasso package: A memory- and computation-efficient solver for
  lasso model fitting with big data in r.
\newblock \emph{R Journal}, 12\penalty0 (2):\penalty0 6--19, 2021.
\newblock URL \url{https://doi.org/10.32614/RJ-2021-001}.

\bibitem[Zhang(2010)]{Zhang2010}
C.~H. Zhang.
\newblock Nearly unbiased variable selection under minimax concave penalty.
\newblock \emph{Annals of Statistics}, 38\penalty0 (2):\penalty0 894--942,
  2010.

\bibitem[Zhou et~al.(2018)Zhou, Nielsen, Fritsche, Dey, Gabrielsen, Wolford,
  LeFaive, VandeHaar, Gagliano, Gifford, et~al.]{Zhou2018}
Wei Zhou, Jonas~B Nielsen, Lars~G Fritsche, Rounak Dey, Maiken~E Gabrielsen,
  Brooke~N Wolford, Jonathon LeFaive, Peter VandeHaar, Sarah~A Gagliano, Aliya
  Gifford, et~al.
\newblock Efficiently controlling for case-control imbalance and sample
  relatedness in large-scale genetic association studies.
\newblock \emph{Nature genetics}, 50\penalty0 (9):\penalty0 1335--1341, 2018.

\end{thebibliography}

\appendix
\section{Supplemental Material}

\subsection{Availability of data and materials}

The \textbf{plmmr} package has been published on GitHub and made available on CRAN at \url{https://cran.r-project.org/web/packages/plmmr/index.html}. The package ships with three example datasets, one for each type of input: (1) data that is read into memory, (2) delimited file input, and (3) a set of PLINK files (.bed/.bim/.fam) input. The documentation website (\url{https://pbreheny.github.io/plmmr/}).
includes tutorial-style articles with hands-on examples of how to analyze data from each of these formats. Users are able to work through the examples in these articles interactively using the datasets that are included with \textbf{plmmr} installation.

All of the code presented in relation to the PennCath data example (as described in Section \ref{sec:penncath} below) has been made available in a public GitHub repository: (\url{https://github.com/tabpeter/demo_plmmr}). This public repository includes a link to the download for the published GWAS data, so that readers may download the PennCath GWAS data, clone the repository, install \textbf{plmmr}, and then reproduce the figures shown here on their own machines.

The GWAS data from the Pittsburgh Orofacial Cleft study are hosted on dbGaP at \url{https://www.ncbi.nlm.nih.gov/projects/gap/cgi-bin/study.cgi?study_id=phs000774.v2.p1}; while we cannot provide access to these protected data, any of our programming scripts may be made available upon request. All analyses were done in R version 4.4.1 \citep{R}.

\subsection{List of abbreviations}

\begin{itemize}
    \item GRM: genomic relatedness matrix
    \item GWAS: genome-wide association study 
    \item LMM: linear mixed model 
    \item PLMM: penalized linear mixed model 
    \item POFC: Pittsburgh Orofacial Cleft Studies 
    \item SNP: single nucleotide polymorphism
\end{itemize}

\subsection{Quality control procedures for PennCath data}


The quality control steps implemented for the PennCath data were as follows: 
\begin{enumerate}
    \item All variants with missing call rates exceeding 0.1 were excluded from analysis.
    \item All variants which had a Hardy-Weinberg equilibrium exact test p-value below 1e-10 were excluded from analysis. 
    \item Variants with a minor allele frequency (MAF) below 0.01 were excluded from analysis. 
    \item Samples with missing call rates exceeding 0.1 were excluded from the analysis. 
\end{enumerate}
All quality control (QC) was done in PLINK v. 1.9 \citep{plink}.

\subsection{Quality control procedures for POFC data}

Details about QC for the samples:  
\begin{itemize}
    \item raw PLINK data: $N = 11,855$
    \item  2 samples removed for high degree of missingess ($> 0.05$ of variants missing)
    \item 3 samples removed for sex discrepancies 
    \item 1,305 samples removed due to not having complete data in corresponding phenotype file 
    \item analytical sample: $N = 10,545$
\end{itemize}
Details about QC for variants: 
\begin{itemize}
    \item 512,926 autosomal variants passed QC filters (same as those for PennCath data)
    \item 469,577 variants had a MAF $> 0.0001$; these were the variants in our analysis. 
\end{itemize}

\subsection{Comparison of plmmr and PenalizedGLMM runtime}

\end{document}